\documentclass[multi]{cambridge6Atight}
\usepackage{natbib}

\usepackage{rotating}
\usepackage{floatpag}
\rotfloatpagestyle{empty}

\usepackage{amsmath}
\usepackage{amsthm}
\usepackage{graphicx}
\usepackage{mathptmx}
\usepackage{xspace}
\usepackage{aas_macros}

  \usepackage{makeidx}
  \makeindex

\copyrightline{This material has been published in \textit{The Impact of Binaries on Stellar Evolution}, Beccari G. \& Boffin H.M.J. (Eds.).
This version is free to view and download for personal use only. Not for re-distribution, re-sale or use in derivative works. \copyright\ 2018 Cambridge University Press.}

\newcommand{\msun}{\mbox{${\rm M}_{\odot}$}\xspace}

\begin{document}


  \alphafootnotes
   \author[G. Nelemans]
    {Gijs Nelemans}

  \chapter{Binaries as Sources of Gravitational Waves}

  \contributor{Gijs Nelemans
    \affiliation{Department of Astrophysics/IMAPP, Radboud University,
      Nijmegen, The Netherlands\newline Institute of Astronomy, KU
      Leuven, Leuven, Belgium}}

 \begin{abstract}
With the discovery of both binary black hole mergers and a binary
neutron star merger the field of Gravitational Wave Astrophysics has
really begun. The current advanced LIGO and Virgo detectors are laser
interferometers that will improve their sensitivity in the coming
years. In the long run, new detectors such as LISA and the Einstein
Telescope will have sensitivities that allow the detection of many
thousands of sources and ET can observe essentially the whole
observable Universe, for heavy black holes. All these measurements
will provide new answers to open questions in binary evolution, related
to mass transfer, out-of-equilibrium stars and the role of
metallicity. In addition, the data will give new constraints on
uncertainties in the evolution of (massive) stars, such as stellar
winds, the role of rotation and the final collapse to a neutron star
or black hole. For black hole binaries, the number of detections is
rapidly approaching 10 and the first proper statistical studies of the
population can be done soon. In the long run, the thousands of
detections by ET will enable us to probe their population in great
detail over the history of the Universe. For neutron stars, the first
question is whether the first detection GW170817 is a typical source
or not. In any case, it has spectacularly shown the promise of
complementary electro-magnetic follow-up. For white dwarfs we have to
wait for LISA (around 2034) but new detections by e.g. Gaia and LSST
will prepare for the astrophysical exploitation of the LISA
measurements.
 \end{abstract}

\section{Gravitational Waves and their Detection}
\label{NelemansGWandDetection}

\index{gravitational waves}With the discovery of first the binary
black hole merger GW150914\index{GW150914} \citep{2016PhRvL.116f1102A}
and recently the binary neutron star merger GW170817\index{GW170817}
\citep{2017PhRvL.119p1101A,2017ApJ...848L..12A}, the field of
Gravitational Wave (GW) astrophysics has been opened in a spectacular
fashion.

Since the original prediction of GW
\citep{1916SPAW.......688E,1918SPAW.......154E}, the field has made a
long journey. Since the 1960s, the importance of GW for binary stars
and the importance of binaries as GW sources has been recognised
\citep[e.g.][]{pac67, pt72}. Since the 1970s, binary stars, in
particular those with neutron star (NS), white dwarf (WD) and
(hypothetical) black hole (BH) components, have been considered the
most promising sources for detectable GW
\citep[e.g.][]{1979A&A....72..120C,ty79a,nyp01,bkb02}. With the
development of laser interferometer detectors such as LIGO\index{LIGO} and Virgo\index{Virgo},
the field gradually evolved to the stage were real detections of GW on
Earth could be expected. At the same time, a long history of proposed
GW detectors in space has culminated in the selection of the LISA\index{LISA}
mission as the first space detector \citep{2017arXiv170200786A}.

For excellent reviews about GW and GW sources I refer e.g. to
\citet{1989CQGra...6.1761S,2009LRR....12....2S}. For binaries, the
effect of GW emission is a decrease in orbital energy and angular
momentum \citep[e.g.][]{1964PhRv..136.1224P} leading, in general, to a
shortening of the orbital period and circularization of the
orbit. This will stop once the stars start to interact which, in most
cases, leads to merger of the two stars. Far from the sources, the
effect of GW is a minuscule effect on the metric, that can in
principle be measured.

\subsection{Detectors: Status and Future}
\label{NelemansDetectors}

The way the minute ripples in space time can be measured is
essentially by determining the time difference that a light ray spends
traversing a ``fixed'' distance using laser interferometry\index{laser
  interferometry} \citep{Weiss:1972}. This clever idea, and the skills
and endurance to turn the idea into a working detector was awarded the
2017 Nobel Prize in Physics for Weiss, Thorne and Barish. GW come with
different frequencies/wavelengths (depending on the size of the
source) and detectors need to be properly sized to be sensitive to
certain waves (very roughly source size $\approx$ wavelength $\approx$
detector size).

The currently operational \textbf{ground based laser interferometers}
\textit{advanced LIGO}\index{LIGO}
\citep{2015CQGra..32g4001L,2016PhRvL.116m1103A} and \textit{advanced
  Virgo}\index{Virgo} \citep{2015CQGra..32b4001A} are 4 and 3 km long
vacuum instruments in which laser beams are first accurately
stabilised (both in frequency and in power) and then split over the
long arms. In the arms the lasers are reflected hundreds of times in
Fabry-Perot cavities, increasing the effective size of the detector,
before being combined again and tuned in such a way that almost
perfect destructive interference is achieved. Since this means most of
the light is reflected back towards the laser, a power-recycling
mirror is placed in the path of the laser, reflecting the light back
into the interferometer, vastly increasing the power of the light
stored in the interferometer and thus increasing the sensitivity of
the instrument. This is needed because the length changes are so small
that they are an incredibly small fraction ($\sim10^{-10}$) of the
$1\mu$ wavelength of the laser and thus lead to only tiny changes in
the interferometer power output.

In addition to achieving the necessary accuracy of the measurement,
the detectors need to be isolated from environmental disturbances and
carefully monitored to determine any external noise
components. Seismic noise is the most important noise source at low
frequencies and is suppressed by many orders of magnitude via seismic
isolation mechanisms. After all these measures, there are still large
numbers of transient noise fluctuations that can mimic real GW
signals, collectively know as triggers. By searching for coincident
and identical triggers in multiple detectors, the trigger background
can be very strongly reduced. However, even then the number of triggers
increases very sharply when going to weaker triggers, so there is a
fairly clean separation between noise and real signals \citep[see e.g. fig 4
of][]{2016PhRvL.116f1102A}.

The two \textit{advanced LIGO} and the \textit{advanced Virgo}
detectors are currently offline to undergo further improvement to
increase the sensitivity. The plan is for the detectors to come online
again for the year-long third observing run (O3) about a year after
the end of O2 \citep{2016LRR....19....1A}, i.e. in the fall of
2018/beginning of 2019. After that, further improvements are expected
before the detectors reach their design sensitivity. Additional
detectors, KAGRA\index{KAGRA} in Japan \citep{2013PhRvD..88d3007A} and
LIGO India\index{LIGO India} \citep{M1100296} are expected to join the
GW detector network in the early 2020s. Further improvements on the
sites of the current detectors are envisioned in the later 2020s
(termed A+ and LIGO Voyager for LIGO sites).

On a longer time scale a third generation of detectors is planned,
such as the European Einstein Telescope\index{Einstein Telescope} or
the US Cosmic Explorer\index{Cosmic Explorer}, each with significantly
larger arms (10-40km), possibly underground to reduce the
environmental noise. The sensitivities of these detectors would be at
least a factor 10 better than the advanced detectors
\citep[e.g.][]{2012CQGra..29l4013S}. Boosted by the success of the
second generation detectors, the preparations for the third generation
detectors is taking off\footnote{\tt
  https://gwic.ligo.org/3Gsubcomm/}.

For larger GW wavelengths the \textbf{ESA LISA mission}\index{LISA} is
planned for a launch in 2034 \citep{2017arXiv170200786A} . It consists
of three identical space craft that exchange laser beams in an
interferometer with arms lengths of 2.5 million km. Although the basic
measurement idea is the same as for the ground based detectors, it is
not possible to actually reflect the laser beam back along the arms,
since the divergence of the laser beam would lead to a signal that is
much too weak. Instead the technique of Time Delay Interferometry
\citep{1999ApJ...527..814A} is used, in which at each time and in each
of the space-craft the phase information of the incoming laser beam is
measured by comparing it to a local reference. All this information is
later combined into a virtual interferometer by adding up signals with
the appropriate time shifts to determine the interference that \emph{would
have} resulted if the light had actually followed that path and would
have been physically combined. The advantage of this technique is that
the laser noise largely cancels and that different types of
virtual interferometers can be constructed (referred to as different
TDI configurations/variables). The LISA mission is currently studied
by ESA with a significant contribution from NASA and the final
adoption of the mission is expected in the early 2020s.



\subsection{Order of Magnitude Estimates: Detection Horizons}
\label{NelemansHorizons}

Gravitational waves\index{gravitational waves} are produced by the
sources with a changing quadrupole moment and thus binary stars are
about the most optimal sources for GW generation
\citep[e.g.][]{tho97}. There is a nice paradox that GW are
energetically at the same time the weakest and the strongest phenomena
in the Universe. The GW luminosity of a binary GW source is given by
\citep{pm63}
$$
L_{\rm GW} = \frac{32 G^4}{5 c^5} \frac{M_1^2 M_2^2 (M_1 +
  M_2)}{a^5} 
$$ which due to the extremely small pre factor $G^4/c^5 \sim 10^{-81}$
vanishes almost completely for any human-made experimental set-up and
even for normal binary stars is very small (e.g. the GW luminosity of Algol
is $\sim10^{-6}$ times the solar luminosity). On the other hand
(forgetting factors of a few difference between $M_1, M_2, M_1 + M_2$),
the same equation can be written as
$$
L_{\rm GW} \approx \frac{32 c^5}{5 G} \left(\frac{G M}{c^2 a} \right)^5
$$ which can approach $c^5/G \sim 10^{59}$ erg/s, for systems with $v
\approx c$, i.e. $GM/a \approx c^2$ can thus can reach above $10^{26}$
solar luminosities! This condition ($v \sim c$) is of course quite
extreme, but for binary BHs that are about to merge can be
approached.

The frequency of the emitted GW is simply related to the orbital
frequency $f_{\rm GW} = 2 f_{\rm orb} \sim GM/a^3$. For merging black
holes, the separation at merger scales with the horizon size and thus
mass ($a \sim R \sim M$), i.e. $f_{\rm merge} \sim 1/M^2$: more massive
BHs merge at lower frequencies.

\begin{figure}
  \includegraphics[width=0.8\textwidth]{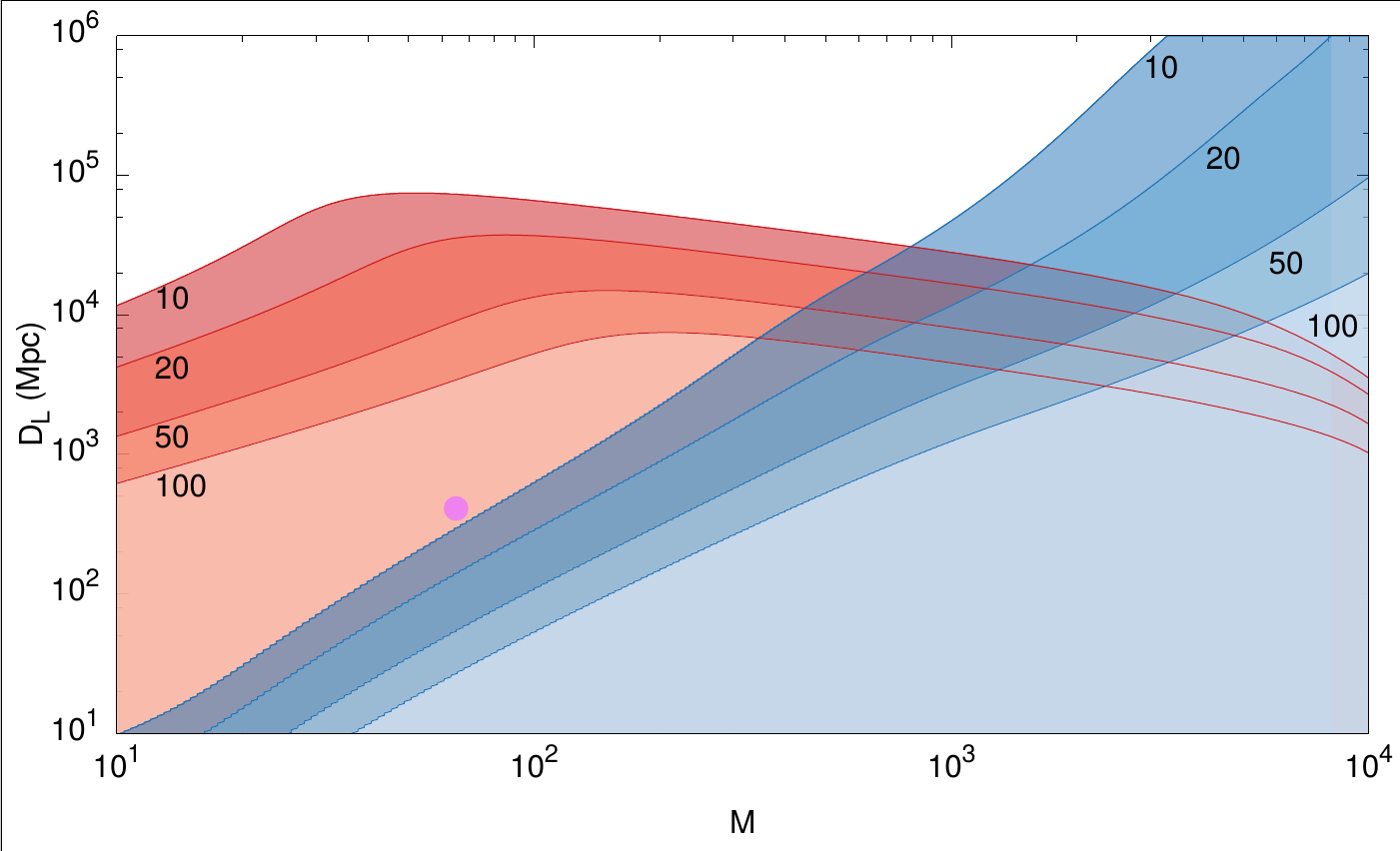}
    \caption[Detection  horizons for  future GW  detectors] {Detection
      horizons as s function of total mass for equal mass binaries for
      future   detectors.   The  expected   S/N   is   given  by   the
      contours.   GW150914\index{GW150914}   is   indicated   by   the
      dot. Figure courtesy Neil Cornish}
    \label{NelemansFig:horizons}
\end{figure}

The (instantaneous) amplitude $h$ of the GW scales with $\sqrt{L_{\rm
    GW})/(f_{\rm GW} d)}$ and thus $h \propto \mathcal{M}^{5/3} f_{\rm
  GW}^{2/3} d^{-1}$, with ``chirp mass'' $\mathcal{M} = (M_1
M_2)^{2/5}/(M_1+M_2)^{1/5}$ and $d$ the distance. We thus can
immediately see that for a binary that is not changing its frequency
during the observation, the signal scales with the (chirp) mass to the
power 5/3, while for such a source the signal becomes stronger for
higher frequency. For systems that do evolve, such as the observed
LIGO/Virgo sources, this picture is not correct, since there is a
limited number of GW cycles observed, which is inversely proportional
to $\mathcal{M}^{5/3}$. The signal strength increases with the square
root of the number of cycles, so scales with $\mathcal{M}^{5/6}$.

In Fig.~\ref{NelemansFig:horizons} the distance to which binaries can
be observed with ET\index{Einstein Telescope} and LISA\index{LISA} is
shown as a function of total mass (assuming equal mass binaries).  The
plot roughly agrees with the scalings derived above, except for masses
above several tens of solar masses in ET, where the signals slowly
move out of the ET sensitivity band (and the maximum distance thus
starts to go down again with mass). Because GW get redshifted, the
maximum sensitivity corresponds to increasingly lower mass systems at
larger distances. From this figure, the enormous promise of GW
astrophysics is immediately clear, since BH binaries can be observed
essentially throughout the whole universe with these instruments!

\section{What can we Learn about Stars and Binaries from GW Measurements?}
\label{NelemansLearn}

For many people, GW detections are primarily interesting from the point
of view of testing General Relativity \citep{2016PhRvL.116v1101A},
having independent cosmological measurements
\citep{2017Natur.551...85A} or detecting signals from the Early
Universe or from unknown sources. For us, GW measurements are a new
way to study binary systems with compact objects, for which there are
many open questions. After listing the most important open questions,
I will discuss what we know and can learn about these questions from GW
observations.

\subsection{Open Questions about Compact Objects}
\label{NelemansOpenQuestions}

Compact objects form at the end of the lives of stars. In this way,
stellar evolution and compact object formation are intimately
linked. If we want to understand the formation of NSs and BHs, we have
to know how the massive stars evolve from which they form
\citep[e.g.][and chapter 9-11 of this
  volume]{2016ApJ...821...38S}. The most pressing issues are stellar
winds and the influence of rotation on the internal evolution of the
stars. But even if these were known, it is not clear how the final
state of the star translates into the formation of the compact
object. The way the core collapses and the possible ensuing supernova
(SN) explosion are still very unclear as are the resulting masses and
spins of the compact objects \citep[e.g.][]{2012ApJ...757...69U}. From
NS observations it is clear that the collapse leads to an asymmetric
kick\citep[e.g.][]{ll94}. The open question is if this also holds for
BHs\citep[e.g.][]{2017MNRAS.467..298R}. All of this is further
complicated by binary interactions that lead to mass transfer and
influence the evolution of the stars. In addition, the binary
evolution processes influence the further evolution of the system and
are typically rather poorly understood (mass transfer and its
resulting loss of mass and angular momentum from the system, including
common-envelope evolution, e.g. chapters 1, 4, 7, 8, 12, 13 of this
volume). Furthermore, if we want to understand compact objects and
binaries in the global context of the evolution of galaxies including
the Milky Way, we also have to understand all these processes in the
changing environment, in particular metallicity. With the new tool of
studying compact objects in binaries with GW, we can hope to shed a
new light on the combined effect of all the uncertainties and thus
learn about them.


\subsection{Black Holes}
\label{NelemansBH}

\begin{figure}
  \includegraphics[width=0.8\textwidth]{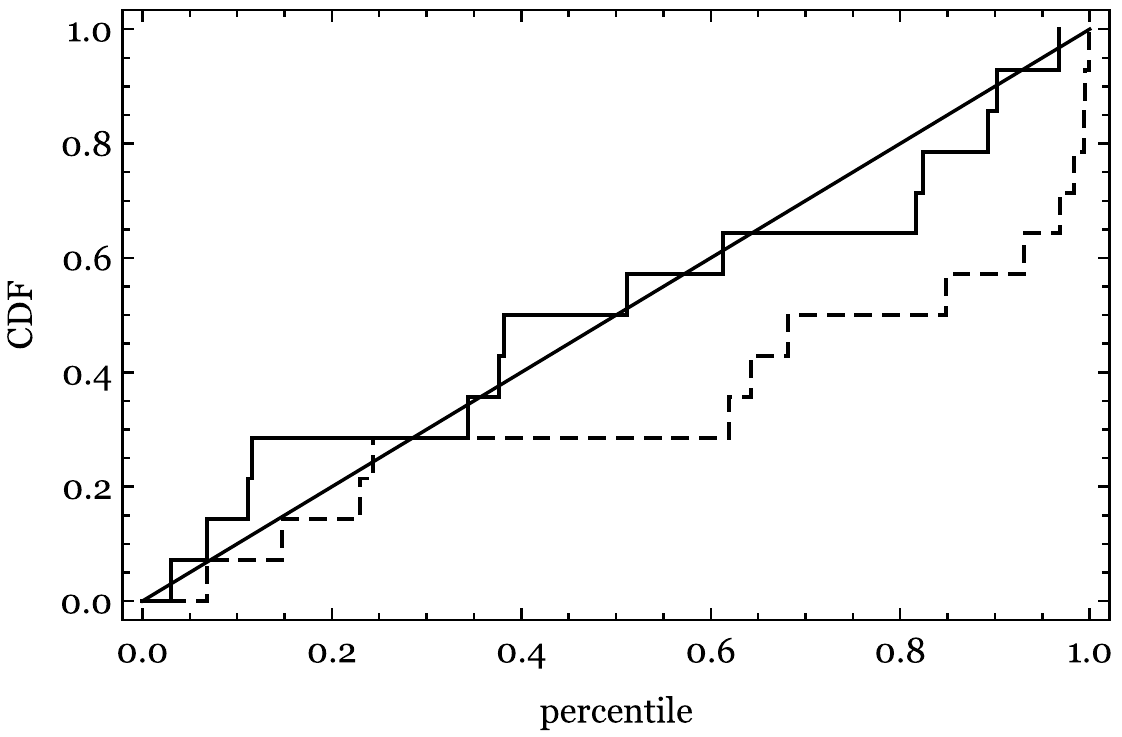}
    \caption[Cumulative distribution of percentiles]{Cumulative
      distribution of the local height above the plane of black-hole
      X-ray binaries compared to a simulated population that
      originates in the disk with low (dashed) and high (solid)
      kicks. From \citet{2017MNRAS.467..298R}}
    \label{NelemansFig:BHkick}
\end{figure}

BHs are the most enigmatic objects, not only by their nature,
but also in their formation. We simply do not know which stars form
BHs and if this happens via initial formation of a neutron
star that later collapses or not and how often (if at all) BH
formation is accompanied by a SN
\citep[e.g.][]{2012ApJ...757...69U}.

A particularly important question is the existence of asymmetric kicks
at the formation of BHs. Several individual black-hole X-ray binaries
have been studied in detail in order to investigate if proof can be
given that BH kicks\index{black hole kicks} exist
\citep[e.g.][]{nth99,whl+04,2009ApJ...697.1057F}. Although kicks are
possible or even plausible, there is not unambiguous evidence. An
alternative approach is to look at the positions of the observed
black-hole X-ray binaries in the Milky Way and compare them to those
of neutron-star X-ray binaries
\citep{wp96,jn04,2012MNRAS.425.2799R}. The more recent of these
studies show that the black-hole systems are found at similar height
above the plane as NS systems or at least surprisingly high up if BHs
do not receive significant ($\sim 100$km/s) kicks. However, NS systems
may have formed in a different way. Also, the observed black-hole systems
are differently biased because the companion stars have to be studied
in detail in order to show that the compact object is so massive that
it has to be a BH. In a systematic study all these effects are taken
into account and the local height of BH X-ray binaries is compared to
a simulated population for different kicks and rescaled to the local
Galactic potential such that all systems can be compared
\citep{2015MNRAS.453.3341R, 2017MNRAS.467..298R}. The result is shown
in Fig.~\ref{NelemansFig:BHkick} and suggests that indeed at least
some BH get relatively high kicks.

\begin{figure}
  \includegraphics[width=0.8\textwidth]{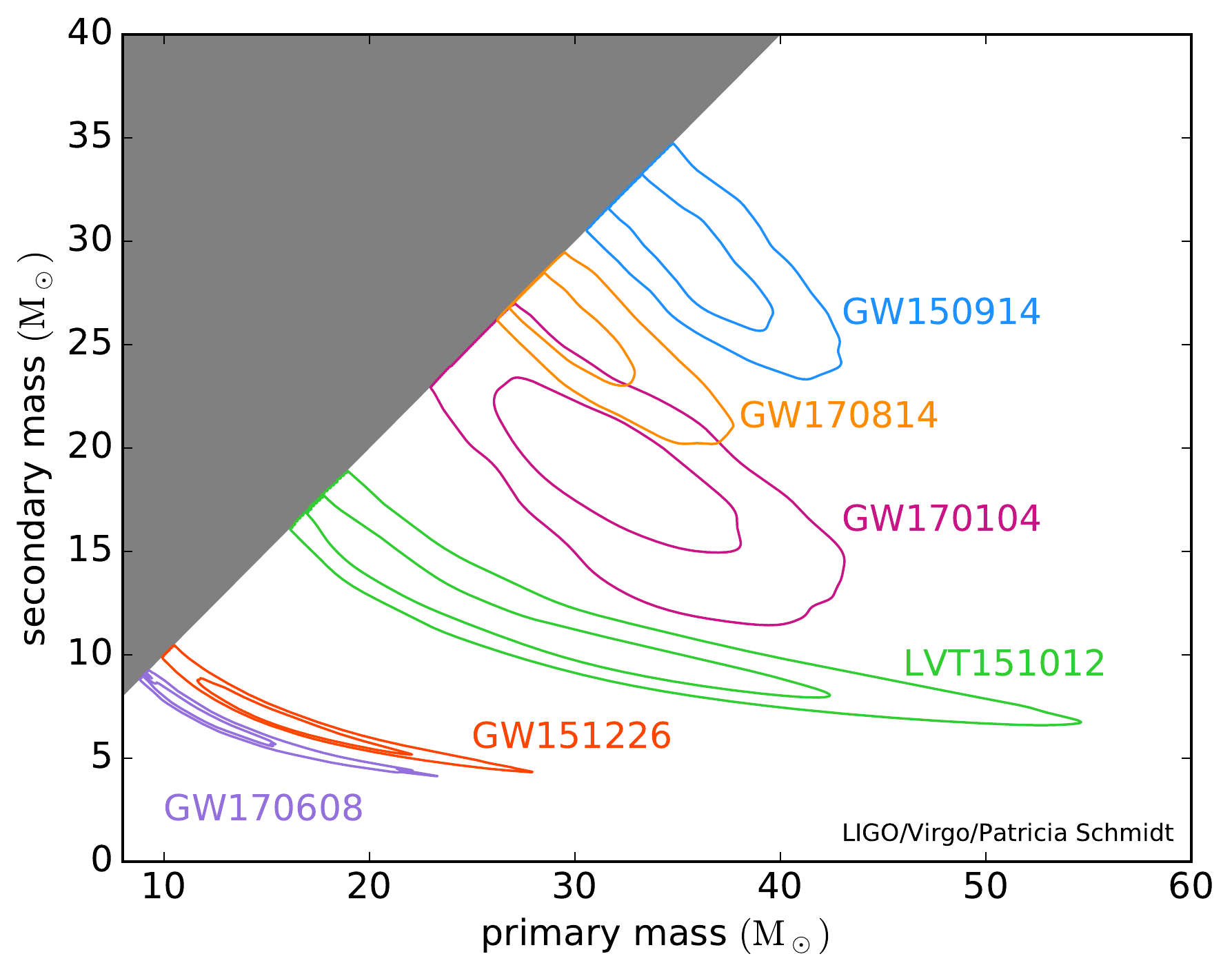}
    \caption[Masses of detected double BHs] {Mass measurements
      of the current binary BH GW detections. Courtesy
      LIGO/Virgo collaboration/Patricia Schmidt}
    \label{NelemansFig:BBH}
\end{figure}

At the same time, simulations of the formation of binary BH as GW
sources suggest that (too) large kicks disrupt too many systems,
leading to a very low merger rate
\citep[e.g.][]{2016ApJ...819..108B}. The LIGO/Virgo detections have
spectacularly opened a new way to (statistically) study the BH
population. Including the most recent announced detection,
GW170608\index{GW170608}, five high-confidence detections and one
candidate (LVT151012) detection have been reported
\citep{2016PhRvX...6d1015A,2017arXiv171105578T,2017PhRvL.118v1101A,2017PhRvL.119n1101A}. They
span a wide range of masses (Fig.~\ref{NelemansFig:BBH})\index{black
  hole masses} ranging from ``heavy'' BH with masses over 30 \msun to
ones in the range of masses known from X-ray binaries
\citep{2014SSRv..183..223C}. The rapid increase in sources makes for
an exiting expectation that soon we will be able to put real
constraints on the mass distribution of binary BH. Looking at
Fig.~\ref{NelemansFig:BBH} it will be the question whether there is a
bi-modal distribution with low and high mass systems, or a continuum,
especially with the uncertain character of the intermediate system
LVT151012.

\begin{figure}
  \includegraphics[width=0.8\textwidth]{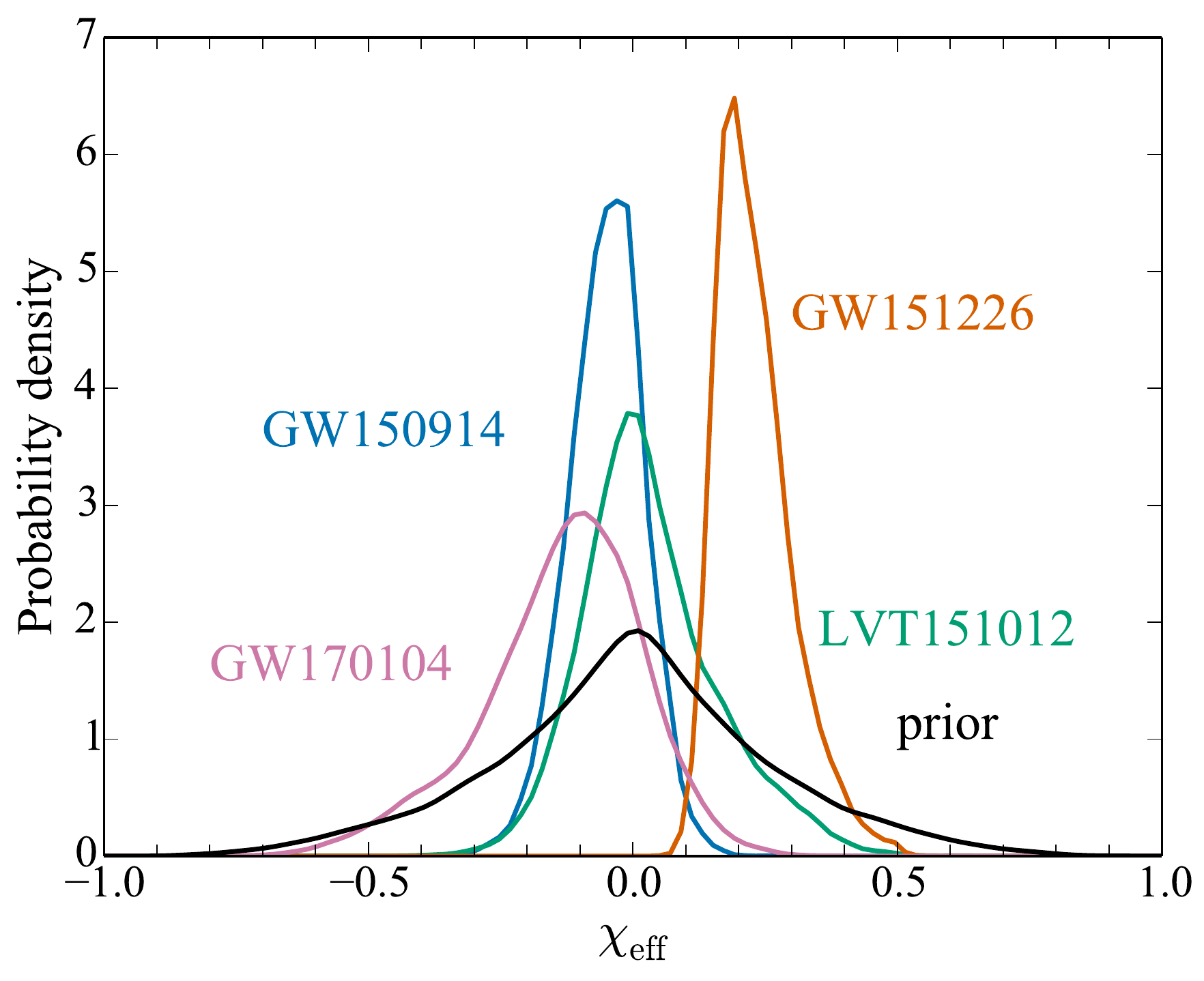}
    \caption[Effective spins of detected binary BH] {Effective spin
      measurements of the current binary BH GW
      detections. From \citet{2017PhRvL.118v1101A}}
    \label{NelemansFig:eff_spin}
\end{figure}

Apart from the masses, the observations have made it possible to
estimate the binary BH merger rate\index{black hole merger rate}. All
detections, including the newer ones are consistent with the latest
rate determination \citep{2017PhRvL.118v1101A} of $12 - 213$
/yr/Gpc$^{3}$. The last piece of information that is available from
the data is en estimate of the ``effective spin'', i.e. the mass
weighted component of the spin parallel to the orbital angular
momentum, see Fig.~\ref{NelemansFig:eff_spin}. As can be seen, most of
the systems have effective spins close to zero.

In our current understanding, there are several ways in which binary
BH that merge can form (apart from ``exotic'' scenarios such as
primordial BH, pop III stars etc) see \citet{2016ApJ...818L..22A}. The
first is the same path that has been put forward for the double NS
systems found in the 1970s \citep[e.g.][]{fh75} but then for more
massive stars
\citep[e.g.][]{2016ApJ...819..108B,2016Natur.534..512B}. There are
several variations on this scenario, in particular avoiding the
common-envelope stage, either by invoking stable mass transfer,
possible due to the large masses of the BH \citep{2017MNRAS.471.4256V}
or due to the rapid rotation of the massive stars, leading to
so-called homogeneous evolution, in which expansion to giant
dimensions is prevented by enhanced mixing
\citep{2016MNRAS.460.3545D}. A fundamentally different formation
scenario is via dynamical interactions in dense stellar environments
\citep{1993Natur.364..423S,pm00} that is found to be a viable
mechanism to produce mergers such as GW150914\index{GW150914}
\citep[e.g.][]{2016PhRvD..93h4029R,2016MNRAS.459.3432M,2017MNRAS.469.4665P}. There
is a lot of excitement about the idea that the effective spins will be
the key to distinguishing different formation channels or inferring BH
kicks. The idea is that for dynamically formed systems the angles
between the BH spins would be random, while for binaries they would be
more or less aligned, unless the BH get very large kicks. However, as
long as there is no clear indication that the BH spin is directly
related to the spin of the progenitor star these considerations are
not very convincing. The reason to doubt the relation between
progenitor spin and BH spin is the observed very large misalignment
between the spins in the double pulsar \citep{2008Sci...321..104B} .

Another extremely nice and promising result is the realisation that
GW150914-like sources are also promising LISA\index{LISA} sources
\citep{2016PhRvL.116w1102S}. Apart from a multitude of nice
measurements when combining LISA and ground based measurements, there
is some chance to detect any remaining eccentricity in the orbits in
the LISA band, that would point towards dynamical formation
\citep[e.g.][]{2016ApJ...830L..18B}.

It is clear that in the next years we will have a steady increase in
the number of detected binary BHs. With improved detector sensitivity,
first with the current second generation of detectors and of course
more spectacular with the third generation detectors, the numbers
could start to add up to very significant numbers. This will allow
fantastic studies. My personal expectation is that with several tens
of detections we we will get a more solid view of the mass and spin
distributions, maybe seeing the first evidence for different
populations if they exist. With several hundreds of detections we can
start to bin up the measurements, for instance in redshift bins and
study the evolution of the system properties. At the same time, the
overall mass and spin distributions, including e.g. mass gaps, upper
and lower limits to BH masses etc., will be determined quite
accurately. Finally, for the third generation detectors, with
thousands and thousands of detections, we can really determine all
these properties as a function of the evolution of the universe!



\subsection{Neutron Stars and Electromagnetic Follow-Up}
\label{NelemansNS}

Since the discovery of PSR J1913+16\index{PSR J1913+16} \citep{ht75},
and maybe even before, double NS have been the prime example GW
sources \citep[see e.g.][]{phi91}. The discovery of the orbital change due
to GW \citep{tw82} started a process that recently reached a
spectacular high light with the discovery of the first double NS GW
source GW170817\index{GW170817}, that also has been detected in
gamma-rays, optical/infrared, X-rays and radio emission
\citep[see][]{2017ApJ...848L..12A,2017PhRvL.119p1101A}.

The current knowledge about the 15 Galactic double NS\index{double
  neutron star} is nicely summarised in
\citet{2017ApJ...846..170T}. From these systems, it is clear that
there is a sizable population of double NS that will merge in the
Galaxy. The source properties, in particular the fact that at short
orbital period there are systems with low and with high
eccentricities, suggest that there may be more than one way to form
them. It has been suggested, that the low eccentricities arise from NS
with small kicks, formed in electron-capture SN
\citep{2004ApJ...612.1044P} that occur much more frequently in close
binary systems than in single stars or wide binaries. However, the
recent finding that even a significant fraction of normal radio
pulsars has low velocities \citep{2017arXiv170808281V} raises some
questions about the one to one link between low kicks and electron
capture SN.

The formation of double NS has been outlined already in the 1970s
\citep{fh75,1975Ap&SS..36..219D,1976Ap&SS..40..115M}. It starts with
two massive stars in which, after a phase of mass transfer, the
primary explodes in a SN. If the orbit remain bound, the system
temporarily takes the form of a High-Mass X-ray binary in which the NS
accretes from the stellar wind of the companion after which it goes
through a second phase of mass transfer that leads to strong spiral in
\citep{1973A&A....25..387V}. The resulting close binary of a helium
burning star and a NS may become an X-ray binary again (like Cyg
X-3\index{Cyg X-3}) after which the helium star explodes to become the
second NS. GW emission subsequently brings the system to merger.

The expectation, based on simulations, that after the merger of two NS,
some of the material is ejected and the possibility that in the merger
(temporarily) a massive, rapidly spinning NS is formed, have led to
the suggestion that the merger of two NS should be accompanied by an
electro-magentic signal. In particular, the radioactive decay of the
ejected low-density NS material, expected to have a brightness in
between novae and SN, is often called kilonova\index{kilonova} or
macronova \citep[see][for a review]{2017LRR....20....3M}. The basic
physics is very elegantly described in \citet{1998ApJ...507L..59L} and
\citet{2005astro.ph.10256K} and shows that the electro-magnetic signal
of the expanding ejecta essentially depends on the injected energy and
the opacity of the material. \citet{2013ApJ...775...18B} showed that
the opacity of the material may be much higher than Thomson
scattering. For still relatively low opacities, the emission comes out
early and fast and is blue; for high opacities the emission comes out
later and slower and is red.

\begin{figure}
  \includegraphics[width=0.8\textwidth]{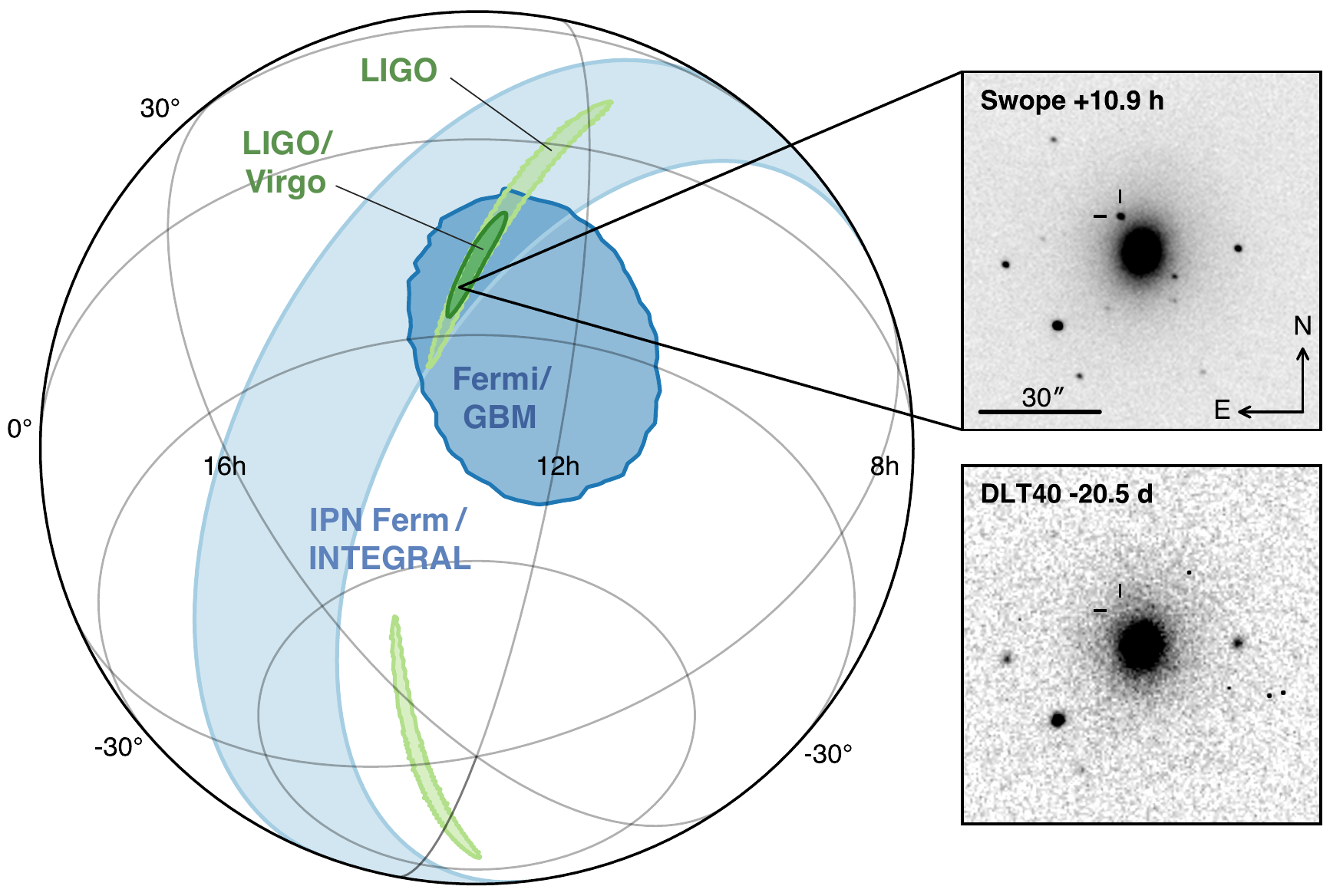}
    \caption[GW170817 counterpart] {Fermi, Integral and LIGO/Virgo
      localisation of the GW170817 and the subsequent detected optical
      counterpart. From \citet{2017ApJ...848L..12A}}
    \label{NelemansFig:GW170817}
\end{figure}

In order to fully exploit the promise of multi-messenger GW astronomy,
the GW signals need to be quickly identified and shared with EM
partners. The experience in the first observing run
\citep{2016ApJ...826L..13A} has been used to improve the procedures
and after the discovery of GW170817\index{GW170817} the alert went out
within 1 hour. In the first day, several tens of observatories have
reacted and observed the counterpart in all bands \citep[see][and
  Fig.~\ref{NelemansFig:GW170817}]{2017ApJ...848L..12A}. The
optical/infrared counterpart is consistent with the predictions of the
kilonova models
\citep{2017Natur.551...75S,2017ApJ...848L..17C,2017Natur.551...67P,2017Sci...358.1570D,2017Sci...358.1559K,2017ApJ...848L..28L}
 although the bright transient implies a quite high ejecta mass
($\sim$0.03 \msun). The masses inferred from the GW detection are
similar to the ones of the know double NS \citep{2017PhRvL.119p1101A}.

Even though it is only one detection, it can serve to make an estimate
of the merger rate of double NS\index{double neutron star merger
  rate}: $1540^{+3200}_{-1220}$/Gpc$^3$/yr. Although within the range
of expectations \citep[e.g.][]{2010CQGra..27q3001A}, it is quite high
compared to the recent theoretical models and not easy to reconcile
with the models without overestimating the binary BH merger rate
\citep{2017arXiv170807885C}. This means that the next observing run
will very quickly show if this single detection was a statistical
outlier, or if the merger rate really is so high.

With this in mind, and the question whether the high ejecta mass and
thus bright kilonova are typical or not, it is important to
systematically investigate the GW error boxes in the future. For that,
for instance the BlackGEM telescope array is developed. It consists of
three 65cm telescopes, each with a 2.7 square degree field of view
that can reach mag 23 in 5 min integration
\citep{2016SPIE.9906E..64B}.

For double NS the expectation is that \textit{LISA} will observe these systems
only in the Milky Way, so expected numbers are low
\citep[e.g.][]{nyp01}. However, a very nice feature is that the
GW strength of the double NS with periods shorter than about 30 min is
such that the measurement of the population in that period range will
be \emph{complete} \citep{2013arXiv1305.5720E}.

Overall, the promise of GW astrophysics for the study of NS binaries
lies mainly in the determination of the rates and properties of the
systems that can be related to the preceding binary evolution. In
particular if also EM signals are detected, the study can be extended
to included detailed studies of the parent population and the ages of
the mergers, probing the delay time between the formation of the
double NS and the merger. 



\subsection{White Dwarfs and Electromagnetic Data}
\label{NelemansWD}

For WD the LIGO and Virgo detectors are too small. Indeed, any ground
based detector will have difficulty detecting any WD, so we have to
turn to LISA. It was realised early on \citep{eis87} that
double WD would form a large population of detectable sources and even
so large that they could form a confusion noise, or foreground
noise \citep[e.g.][]{hbw90}. These systems are now found regularly
optically \citep[e.g.][]{2011CQGra..28i4019M,2017ApJ...847...10B}. With the
current design of the mission \citep{2017arXiv170200786A}, the
sensitivity of LISA is such that with a recent model for the Galactic
population of double WD \citep{2012A&A...546A..70T} the expectation is
that it will detect this foreground, in addition to maybe 25,000
individually resolved systems
\citep{2017JPhCS.840a2024C,2017MNRAS.470.1894K}. A small, but not
negligible fraction will also be detectable with EM instruments
\citep[e.g.][]{2013MNRAS.429.2361L}.

The emission of GW will bring close double WD closer and eventually
leads to Roche-lobe overflow of the lowest-mass WD. Contrary to the
case of two NS, there is a possibility for stable mass transfer
\citep[see][]{mns03}. The expected properties of such interacting
double WD correspond closely to the observed AM CVn systems
\citep[see][]{2010PASP..122.1133S}. These are characterised by very
short orbital periods (5-60 min), helium accretion discs around WD
with and very low-mass, unseen companions that need to be degenerate
in order to fit in the orbits.

\begin{figure}
  \includegraphics[width=0.8\textwidth]{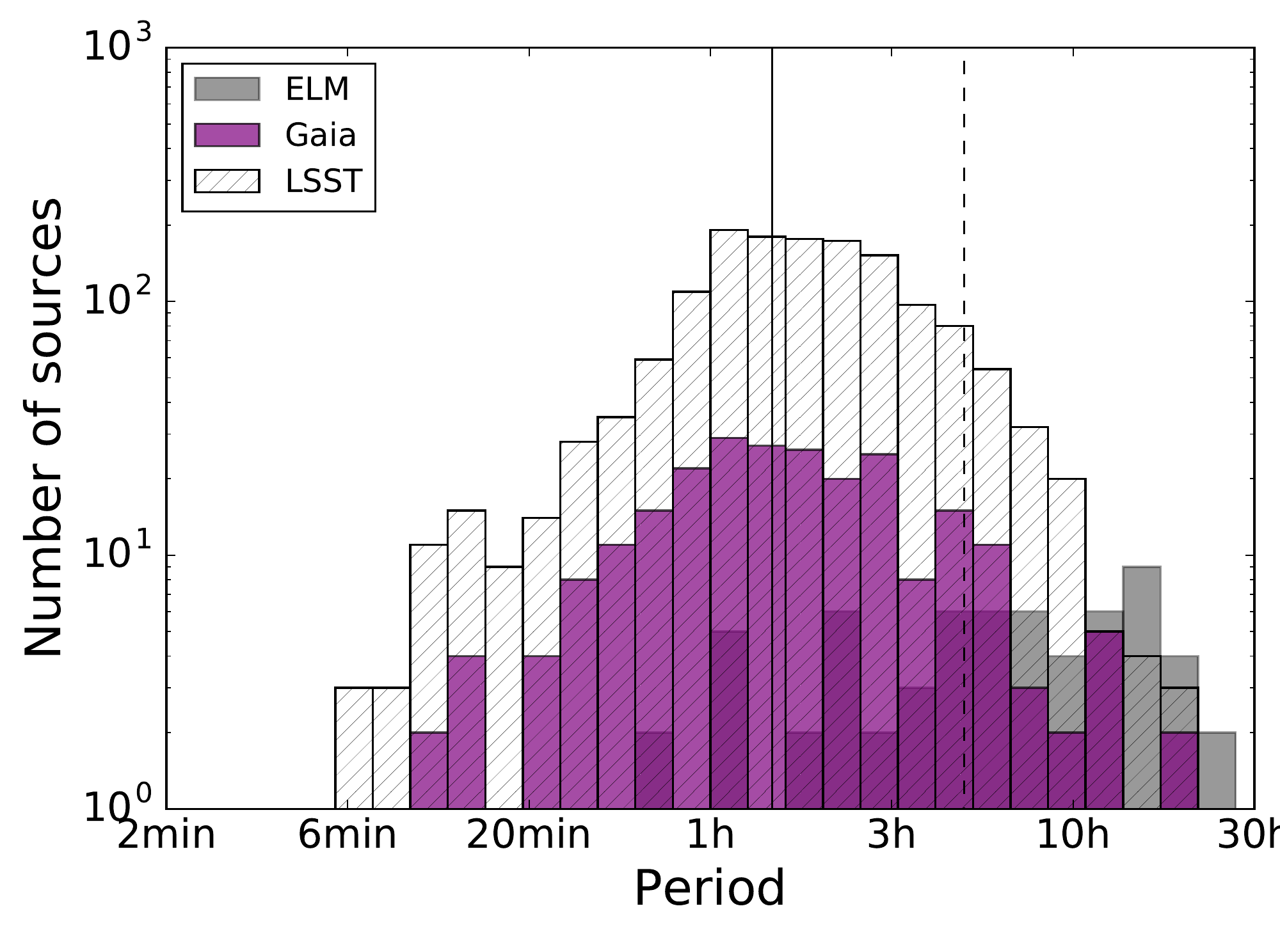}
    \caption[LISA WD sources with EM counterpart] {Expected LISA
      double WD sources for which an optical
      counterpart can be detected by LSST or Gaia. From
      \citet{2017MNRAS.470.1894K}}
    \label{NelemansFig:Korol}
\end{figure}

Some of the observed systems are expected to be individually detected
by LISA and are referred to as verification binaries
\citep[e.g.][]{sv06,2015JPhCS.610a2003S,2017ApJ...847...10B}. Between
now and the launch of LISA there will be different instruments that
have the potential to increase the number of verification binaries
significantly, in particular Gaia and LSST \citep[][and see
  Fig.~\ref{NelemansFig:Korol}]{2017MNRAS.470.1894K,2017arXiv171008370B}. The
combined data from GW and EM observations, in particular period evolution
measurements, e.g. through eclipse timings, have the potential to
greatly improve out understanding of the evolution into double WD and
the small fraction that turns into AM CVn systems
\citep{2014ApJ...790..161S,2014ApJ...791...76S,2017arXiv171008370B}

We can expect to get information about the binary evolution leading to
the formation of double WD and the initial binary populations for
low-mass stars from the numbers and properties of the large sample of
detected systems. In particular when the (chirp) masses can be
measured (only in a subset of the individual systems), constraints on
the common envelope and the stability of mass transfer between two WD
can be inferred. Finally, the systems with measured distances (even
fewer or the systems with complementary EM data) can be used to study
the Galactic distribution of systems and, since some of these will be
located throughout the Galaxy, the structure of the Milky Way \citep[e.g.][]{2018arXiv180603306K}.



\section{Conclusions}
\label{NelemansConclusions}

We now really can say that GW astrophysics has started and is not just
something for the future. At least for the next decade, binaries will
be the key GW sources. With the plans for increased sensitivity, the
prospects are excellent, not only for more detections, but also for
sufficiently large samples that allow statistical studies. For binary
BH we can expect many detections. At the moment, their formation is
unclear and hopefully the statistical studies of the population will
start to give directions on how to tackle the question of their
origin. For double NS the first question is how many detections we
will see in the next years. In addition, the first detection was such
a rich event, with detections in all EM bands, that we will have to
see how typical such an event is. In any case, the study of EM
counterparts will provide valuable extra information that will make it
easier to use them to study binary evolution. Indeed, if the merger
rate is as high as suggested by this first detection, we may already
have to go back to the drawing board for their evolution. In the long
run, detectors such as LISA and ET will start the phase of mass
detections, with thousands of sources expected. The WD binaries detected
by LISA will provide new information on low-mass binary formation and
the delicate process of mass transfer between WD. Before the LISA
launch, many new systems will be detected and long-term monitoring of
(eclipsing) sources will provide crucial information to complement the
LISA measurements. Overall, it is clear that our understanding of
binary evolution will get a boost from all these new data!


\bibliography{binaries}\label{refs}

\begin{thebibliography}{98}
\expandafter\ifx\csname natexlab\endcsname\relax\def\natexlab#1{#1}\fi
\expandafter\ifx\csname selectlanguage\endcsname\relax
  \def\selectlanguage#1{\relax}\fi

\bibitem[\protect\citename{{Aasi} {et~al.}, }2015]{2015CQGra..32g4001L}
{Aasi} J., , {Abbott} B.~P., , {Abbott} R.,    et~al. 2015.
\newblock {Advanced LIGO}.
\newblock {\em Classical and Quantum Gravity}, {\bf 32}(7), 074001.

\bibitem[\protect\citename{{Abadie} {et~al.}, }2010]{2010CQGra..27q3001A}
{Abadie} J., , {Abbott} B.~P., , {Abbott} R.,    et~al. 2010.
\newblock {TOPICAL REVIEW: Predictions for the rates of compact binary
  coalescences observable by ground-based gravitational-wave detectors}.
\newblock {\em Classical and Quantum Gravity}, {\bf 27}(17), 173001.

\bibitem[\protect\citename{{Abbott} {et~al.}, }2016a]{2016ApJ...818L..22A}
{Abbott} B.~P., , {Abbott} R., , {Abbott} T.~D.,    et~al. 2016a.
\newblock {Astrophysical Implications of the Binary Black-hole Merger
  GW150914}.
\newblock {\em \apjl}, {\bf 818}, L22.

\bibitem[\protect\citename{{Abbott} {et~al.}, }2016b]{2016PhRvX...6d1015A}
{Abbott} B.~P., , {Abbott} R., , {Abbott} T.~D.,    et~al. 2016b.
\newblock {Binary Black Hole Mergers in the First Advanced LIGO Observing Run}.
\newblock {\em Physical Review X}, {\bf 6}(4), 041015.

\bibitem[\protect\citename{{Abbott} {et~al.}, }2016c]{2016PhRvL.116m1103A}
{Abbott} B.~P., , {Abbott} R., , {Abbott} T.~D.,    et~al. 2016c.
\newblock {GW150914: The Advanced LIGO Detectors in the Era of First
  Discoveries}.
\newblock {\em Physical Review Letters}, {\bf 116}(13), 131103.

\bibitem[\protect\citename{{Abbott} {et~al.}, }2016d]{2016ApJ...826L..13A}
{Abbott} B.~P., , {Abbott} R., , {Abbott} T.~D.,    et~al. 2016d.
\newblock {Localization and Broadband Follow-up of the Gravitational-wave
  Transient GW150914}.
\newblock {\em \apjl}, {\bf 826}, L13.

\bibitem[\protect\citename{{Abbott} {et~al.}, }2016e]{2016PhRvL.116f1102A}
{Abbott} B.~P., , {Abbott} R., , {Abbott} T.~D.,    et~al. 2016e.
\newblock {Observation of Gravitational Waves from a Binary Black Hole Merger}.
\newblock {\em Physical Review Letters}, {\bf 116}(6), 061102.

\bibitem[\protect\citename{{Abbott} {et~al.}, }2016f]{2016LRR....19....1A}
{Abbott} B.~P., , {Abbott} R., , {Abbott} T.~D.,    et~al. 2016f.
\newblock {Prospects for Observing and Localizing Gravitational-Wave Transients
  with Advanced LIGO and Advanced Virgo}.
\newblock {\em Living Reviews in Relativity}, {\bf 19}, 1.
\newblock arXiv:1304.0670.

\bibitem[\protect\citename{{Abbott} {et~al.}, }2016g]{2016PhRvL.116v1101A}
{Abbott} B.~P., , {Abbott} R., , {Abbott} T.~D.,    et~al. 2016g.
\newblock {Tests of General Relativity with GW150914}.
\newblock {\em Physical Review Letters}, {\bf 116}(22), 221101.

\bibitem[\protect\citename{{Abbott} {et~al.}, }2017a]{2017Natur.551...85A}
{Abbott} B.~P., , {Abbott} R., , {Abbott} T.~D.,    et~al. 2017a.
\newblock {A gravitational-wave standard siren measurement of the Hubble
  constant}.
\newblock {\em \nat}, {\bf 551}, 85--88.

\bibitem[\protect\citename{{Abbott} {et~al.}, }2017b]{2017PhRvL.118v1101A}
{Abbott} B.~P., , {Abbott} R., , {Abbott} T.~D.,    et~al. 2017b.
\newblock {GW170104: Observation of a 50-Solar-Mass Binary Black Hole
  Coalescence at Redshift 0.2}.
\newblock {\em Physical Review Letters}, {\bf 118}(22), 221101.

\bibitem[\protect\citename{{Abbott} {et~al.}, }2017c]{2017arXiv171105578T}
{Abbott} B.~P., , {Abbott} R., , {Abbott} T.~D.,    et~al. 2017c.
\newblock {GW170608: Observation of a 19-solar-mass Binary Black Hole
  Coalescence}.
\newblock {\em ArXiv e-prints}, Nov.

\bibitem[\protect\citename{{Abbott} {et~al.}, }2017d]{2017PhRvL.119n1101A}
{Abbott} B.~P., , {Abbott} R., , {Abbott} T.~D.,    et~al. 2017d.
\newblock {GW170814: A Three-Detector Observation of Gravitational Waves from a
  Binary Black Hole Coalescence}.
\newblock {\em Physical Review Letters}, {\bf 119}(14), 141101.

\bibitem[\protect\citename{{Abbott} {et~al.}, }2017e]{2017PhRvL.119p1101A}
{Abbott} B.~P., , {Abbott} R., , {Abbott} T.~D.,    et~al. 2017e.
\newblock {GW170817: Observation of Gravitational Waves from a Binary Neutron
  Star Inspiral}.
\newblock {\em Physical Review Letters}, {\bf 119}(16), 161101.

\bibitem[\protect\citename{{Abbott} {et~al.}, }2017f]{2017ApJ...848L..12A}
{Abbott} B.~P., , {Abbott} R., , {Abbott} T.~D.,    et~al. 2017f.
\newblock {Multi-messenger Observations of a Binary Neutron Star Merger}.
\newblock {\em \apjl}, {\bf 848}, L12.

\bibitem[\protect\citename{{Acernese} {et~al.}, }2015]{2015CQGra..32b4001A}
{Acernese} F., , {Agathos} M., , {Agatsuma} K.,    et~al. 2015.
\newblock {Advanced Virgo: a second-generation interferometric gravitational
  wave detector}.
\newblock {\em Classical and Quantum Gravity}, {\bf 32}(2), 024001.

\bibitem[\protect\citename{{Amaro-Seoane} {et~al.}, }2017]{2017arXiv170200786A}
{Amaro-Seoane} P., , {Audley} H., , {Babak} S.,    et~al. 2017.
\newblock {Laser Interferometer Space Antenna}.
\newblock {\em ArXiv e-prints}, Feb.

\bibitem[\protect\citename{{Armstrong} {et~al.}, }1999]{1999ApJ...527..814A}
{Armstrong} J.~W., , {Estabrook} F.~B.,    {Tinto} M., . 1999.
\newblock {Time-Delay Interferometry for Space-based Gravitational Wave
  Searches}.
\newblock {\em \apj}, {\bf 527}, 814--826.

\bibitem[\protect\citename{{Aso} {et~al.}, }2013]{2013PhRvD..88d3007A}
{Aso} Y., , {Michimura} Y., , {Somiya} K.,    et~al. 2013.
\newblock {Interferometer design of the KAGRA gravitational wave detector}.
\newblock {\em \prd}, {\bf 88}(4), 043007.

\bibitem[\protect\citename{{Barnes} and {Kasen}, }2013]{2013ApJ...775...18B}
{Barnes} J.,  {Kasen} D., . 2013.
\newblock {Effect of a High Opacity on the Light Curves of Radioactively
  Powered Transients from Compact Object Mergers}.
\newblock {\em \apj}, {\bf 775}, 18.

\bibitem[\protect\citename{{Belczynski} {et~al.}, }2002]{bkb02}
{Belczynski} K., , {Kalogera} V.,    {Bulik} T., . 2002.
\newblock {A Comprehensive Study of Binary Compact Objects as Gravitational
  Wave Sources: Evolutionary Channels, Rates, and Physical Properties}.
\newblock {\em \apj}, {\bf 572}, 407--431.

\bibitem[\protect\citename{{Belczynski} {et~al.}, }2016a]{2016ApJ...819..108B}
{Belczynski} K., , {Repetto} S., , {Holz} D.~E.,    et~al. 2016a.
\newblock {Compact Binary Merger Rates: Comparison with LIGO/Virgo Upper
  Limits}.
\newblock {\em \apj}, {\bf 819}, 108.

\bibitem[\protect\citename{{Belczynski} {et~al.}, }2016b]{2016Natur.534..512B}
{Belczynski} K., , {Holz} D.~E., , {Bulik} T.,    {O'Shaughnessy} R., . 2016b.
\newblock {The first gravitational-wave source from the isolated evolution of
  two stars in the 40-100 solar mass range}.
\newblock {\em \nat}, {\bf 534}, 512--515.

\bibitem[\protect\citename{{Bloemen} {et~al.}, }2016]{2016SPIE.9906E..64B}
{Bloemen} S., , {Groot} P., , {Woudt} P.,    et~al. 2016 (July).
\newblock {MeerLICHT and BlackGEM: custom-built telescopes to detect faint
  optical transients}.
\newblock {Page  990664 of:} {\em Ground-based and Airborne Telescopes VI}.
\newblock \procspie, vol. 9906.

\bibitem[\protect\citename{{Breivik} {et~al.}, }2016]{2016ApJ...830L..18B}
{Breivik} K., , {Rodriguez} C.~L., , {Larson} S.~L., , {Kalogera} V.,
  {Rasio} F.~A., . 2016.
\newblock {Distinguishing between Formation Channels for Binary Black Holes
  with LISA}.
\newblock {\em \apjl}, {\bf 830}, L18.

\bibitem[\protect\citename{{Breivik} {et~al.}, }2017]{2017arXiv171008370B}
{Breivik} K., , {Kremer} K., , {Bueno} M.,    et~al. 2017.
\newblock {Characterizing accreting double white dwarf binaries with LISA and
  Gaia}.
\newblock {\em ArXiv e-prints}, Oct.

\bibitem[\protect\citename{{Breton} {et~al.}, }2008]{2008Sci...321..104B}
{Breton} R.~P., , {Kaspi} V.~M., , {Kramer} M.,    et~al. 2008.
\newblock {Relativistic Spin Precession in the Double Pulsar}.
\newblock {\em Science}, {\bf 321}, 104.

\bibitem[\protect\citename{{Brown} {et~al.}, }2017]{2017ApJ...847...10B}
{Brown} W.~R., , {Kilic} M., , {Kosakowski} A.,    {Gianninas} A., . 2017.
\newblock {Discovery of a Detached, Eclipsing 40 Minute Period Double White
  Dwarf Binary and a Friend: Implications for He+CO White Dwarf Mergers}.
\newblock {\em \apj}, {\bf 847}, 10.

\bibitem[\protect\citename{{Casares} and {Jonker}, }2014]{2014SSRv..183..223C}
{Casares} J.,  {Jonker} P.~G., . 2014.
\newblock {Mass Measurements of Stellar and Intermediate-Mass Black Holes}.
\newblock {\em \ssr}, {\bf 183}, 223--252.

\bibitem[\protect\citename{{Chruslinska} {et~al.}, }2017]{2017arXiv170807885C}
{Chruslinska} M., , {Belczynski} K., , {Klencki} J.,    {Benacquista} M., .
  2017.
\newblock {Double neutron stars: merger rates revisited}.
\newblock {\em ArXiv e-prints}, Aug.

\bibitem[\protect\citename{{Clark} {et~al.}, }1979]{1979A&A....72..120C}
{Clark} J.~P.~A., , {van den Heuvel} E.~P.~J.,    {Sutantyo} W., . 1979.
\newblock {Formation of neutron star binaries and their importance for
  gravitational radiation}.
\newblock {\em \aap}, {\bf 72}, 120--128.

\bibitem[\protect\citename{{Cornish} and {Robson}, }2017]{2017JPhCS.840a2024C}
{Cornish} N.,  {Robson} T., . 2017 (May).
\newblock {Galactic binary science with the new LISA design}.
\newblock {Page  012024 of:} {\em Journal of Physics Conference Series}.
\newblock Journal of Physics Conference Series, vol. 840.

\bibitem[\protect\citename{{Cowperthwaite} {et~al.},
  }2017]{2017ApJ...848L..17C}
{Cowperthwaite} P.~S., , {Berger} E., , {Villar} V.~A.,    et~al. 2017.
\newblock {The Electromagnetic Counterpart of the Binary Neutron Star Merger
  LIGO/Virgo GW170817. II. UV, Optical, and Near-infrared Light Curves and
  Comparison to Kilonova Models}.
\newblock {\em \apjl}, {\bf 848}, L17.

\bibitem[\protect\citename{{De Loore} {et~al.}, }1975]{1975Ap&SS..36..219D}
{De Loore} C., , {De Greve} J.~P.,    {de Cuyper} J.~P., . 1975.
\newblock {Evolution of massive close binaries. II - The POST X-ray binary
  stage: Origin of run-away and binary pulsars}.
\newblock {\em \apss}, {\bf 36}, 219--225.

\bibitem[\protect\citename{{de Mink} and {Mandel}, }2016]{2016MNRAS.460.3545D}
{de Mink} S.~E.,  {Mandel} I., . 2016.
\newblock {The chemically homogeneous evolutionary channel for binary black
  hole mergers: rates and properties of gravitational-wave events detectable by
  advanced LIGO}.
\newblock {\em \mnras}, {\bf 460}, 3545--3553.

\bibitem[\protect\citename{{Drout} {et~al.}, }2017]{2017Sci...358.1570D}
{Drout} M.~R., , {Piro} A.~L., , {Shappee} B.~J.,    et~al. 2017.
\newblock {Light curves of the neutron star merger GW170817/SSS17a:
  Implications for r-process nucleosynthesis}.
\newblock {\em Science}, {\bf 358}, 1570--1574.

\bibitem[\protect\citename{{Einstein}, }1916]{1916SPAW.......688E}
{Einstein} A., . 1916.
\newblock {N{\"a}herungsweise Integration der Feldgleichungen der Gravitation}.
\newblock {\em Sitzungsberichte der K{\"o}niglich Preu{\ss}ischen Akademie der
  Wissenschaften (Berlin), Seite 688-696.}

\bibitem[\protect\citename{{Einstein}, }1918]{1918SPAW.......154E}
{Einstein} A., . 1918.
\newblock {{\"U}ber Gravitationswellen}.
\newblock {\em Sitzungsberichte der K{\"o}niglich Preu{\ss}ischen Akademie der
  Wissenschaften (Berlin), Seite 154-167.}

\bibitem[\protect\citename{{eLISA Consortium} {et~al.},
  }2013]{2013arXiv1305.5720E}
{eLISA Consortium}, {Amaro Seoane} P., , {Aoudia} S.,    et~al. 2013.
\newblock {The Gravitational Universe}.
\newblock {\em ArXiv e-prints}, May.

\bibitem[\protect\citename{Evans {et~al.}, }1987]{eis87}
Evans C.~R., , Iben Jr I.,    Smarr L., . 1987.
\newblock Degenerate dwarf binaries as promising sources of gavitaional
  radiation.
\newblock {\em \apj}, {\bf 323}, 129--139.

\bibitem[\protect\citename{Flannery and van~den Heuvel, }1975]{fh75}
Flannery B.~P.,  van~den Heuvel E. P.~J., . 1975.
\newblock On the origin of the binary pulsar PSR 1913 + 16.
\newblock {\em \aap}, {\bf 39}, 61--67.

\bibitem[\protect\citename{{Fragos} {et~al.}, }2009]{2009ApJ...697.1057F}
{Fragos} T., , {Willems} B., , {Kalogera} V.,    et~al. 2009.
\newblock {Understanding Compact Object Formation and Natal Kicks. II. The Case
  of XTE J1118 + 480}.
\newblock {\em \apj}, {\bf 697}, 1057--1070.

\bibitem[\protect\citename{{Hils} {et~al.}, }1990]{hbw90}
{Hils} D., , Bender P.~L.,    Webbink R.~F., . 1990.
\newblock Gravitational radiation from the Galaxy.
\newblock {\em \apj}, {\bf 360}, 75--94.

\bibitem[\protect\citename{Hulse and Taylor, }1975]{ht75}
Hulse R.~A.,  Taylor J.~H., . 1975.
\newblock Discovery of a pulsar in a binary system.
\newblock {\em \apjl}, {\bf 195}, L51--L53.

\bibitem[\protect\citename{{Iyer} {\em et~al.}, }2011]{M1100296}
{Iyer} B.,  et~al. 2011.
\newblock {\em {LIGO India}}.
\newblock Tech. rept. LIGO-M1100296.
\newblock https://dcc.ligo.org/LIGO-M1100296/public.

\bibitem[\protect\citename{{Jonker} and {Nelemans}, }2004]{jn04}
{Jonker} P.~G.,  {Nelemans} G., . 2004.
\newblock {The distances to Galactic low-mass X-ray binaries: consequences for
  black hole luminosities and kicks}.
\newblock {\em \mnras}, {\bf 354}, 355.

\bibitem[\protect\citename{{Kasliwal} {et~al.}, }2017]{2017Sci...358.1559K}
{Kasliwal} M.~M., , {Nakar} E., , {Singer} L.~P.,    et~al. 2017.
\newblock {Illuminating gravitational waves: A concordant picture of photons
  from a neutron star merger}.
\newblock {\em Science}, {\bf 358}, 1559--1565.

\bibitem[\protect\citename{{Korol} {et~al.}, }2017]{2017MNRAS.470.1894K}
{Korol} V., , {Rossi} E.~M., , {Groot} P.~J.,    et~al. 2017.
\newblock {Prospects for detection of detached double white dwarf binaries with
  Gaia, LSST and LISA}.
\newblock {\em \mnras}, {\bf 470}, 1894--1910.

\bibitem[\protect\citename{{Korol} {et~al.}, }2018]{2018arXiv180603306K}
{Korol} V., , {Rossi} E.~M.,    {Barausse} E., . 2018.
\newblock {A multi-messenger study of the Milky Way's stellar disc and bulge
  with LISA, Gaia and LSST}.
\newblock {\em ArXiv e-prints}, June.

\bibitem[\protect\citename{{Kulkarni}, }2005]{2005astro.ph.10256K}
{Kulkarni} S.~R., . 2005.
\newblock {Modeling Supernova-like Explosions Associated with Gamma-ray Bursts
  with Short Durations}.
\newblock {\em ArXiv Astrophysics e-prints}, Oct.

\bibitem[\protect\citename{{Levan} {et~al.}, }2017]{2017ApJ...848L..28L}
{Levan} A.~J., , {Lyman} J.~D., , {Tanvir} N.~R.,    et~al. 2017.
\newblock {The Environment of the Binary Neutron Star Merger GW170817}.
\newblock {\em \apjl}, {\bf 848}, L28.

\bibitem[\protect\citename{{Li} and {Paczy{\'n}ski},
  }1998]{1998ApJ...507L..59L}
{Li} L.-X.,  {Paczy{\'n}ski} B., . 1998.
\newblock {Transient Events from Neutron Star Mergers}.
\newblock {\em \apjl}, {\bf 507}, L59--L62.

\bibitem[\protect\citename{{Littenberg} {et~al.}, }2013]{2013MNRAS.429.2361L}
{Littenberg} T.~B., , {Larson} S.~L., , {Nelemans} G.,    {Cornish} N.~J., .
  2013.
\newblock {Prospects for observing ultracompact binaries with space-based
  gravitational wave interferometers and optical telescopes}.
\newblock {\em \mnras}, {\bf 429}, 2361--2365.

\bibitem[\protect\citename{Lyne and Lorimer, }1994]{ll94}
Lyne A.~G.,  Lorimer D.~R., . 1994.
\newblock HIGH BIRTH VELOCITIES OF RADIO PULSARS.
\newblock {\em Nat}, {\bf 369}, 127--.

\bibitem[\protect\citename{{Mapelli}, }2016]{2016MNRAS.459.3432M}
{Mapelli} M., . 2016.
\newblock {Massive black hole binaries from runaway collisions: the impact of
  metallicity}.
\newblock {\em \mnras}, {\bf 459}, 3432--3446.

\bibitem[\protect\citename{{Marsh}, }2011]{2011CQGra..28i4019M}
{Marsh} T.~R., . 2011.
\newblock {Double white dwarfs and LISA}.
\newblock {\em Classical and Quantum Gravity}, {\bf 28}(9), 094019.

\bibitem[\protect\citename{{Marsh} {et~al.}, }2003]{mns03}
{Marsh} T.~R., , {Nelemans} G.,    {Steeghs} D., . 2003.
\newblock {Direct Impact Accretors: Evolutionary links between detached and
  semi-detached white dwarfs}.
\newblock {Page  275 of:} de Martino D., , Kalytis R., , Silvotti R.,
  Solheim J.,  (eds), {\em White Dwarfs, Proc. XIII Workshop on White Dwarfs}.
\newblock Kluwer.

\bibitem[\protect\citename{{Massevitch} {et~al.}, }1976]{1976Ap&SS..40..115M}
{Massevitch} A.~G., , {Tutukov} A.~V.,    {Iungelson} L.~R., . 1976.
\newblock {Evolution of massive close binaries and formation of neutron stars
  and black holes}.
\newblock {\em \apss}, {\bf 40}, 115--133.

\bibitem[\protect\citename{{Metzger}, }2017]{2017LRR....20....3M}
{Metzger} B.~D., . 2017.
\newblock {Kilonovae}.
\newblock {\em Living Reviews in Relativity}, {\bf 20}, 3.

\bibitem[\protect\citename{Nelemans {et~al.}, }1999]{nth99}
Nelemans G., , Tauris T.~M.,    van~den Heuvel E. P.~J., . 1999.
\newblock Constraints on mass ejection in black hole formation derived from
  black hole X-ray binaries.
\newblock {\em \aap}, {\bf 352}, L87--L90.

\bibitem[\protect\citename{Nelemans {et~al.}, }2001]{nyp01}
Nelemans G., , Yungelson L.~R.,    Portegies~Zwart S.~F., . 2001.
\newblock The gravitational wave signal from the Galactic disk population of
  binaries containing two compact objects.
\newblock {\em \aap}, {\bf 375}, 890--898.

\bibitem[\protect\citename{Paczy\'nski, }1967]{pac67}
Paczy\'nski B., . 1967.
\newblock ?
\newblock {\em Acta Astron.}, {\bf 17}, 287.

\bibitem[\protect\citename{{Park} {et~al.}, }2017]{2017MNRAS.469.4665P}
{Park} D., , {Kim} C., , {Lee} H.~M., , {Bae} Y.-B.,    {Belczynski} K., .
  2017.
\newblock {Black hole binaries dynamically formed in globular clusters}.
\newblock {\em \mnras}, {\bf 469}, 4665--4674.

\bibitem[\protect\citename{{Peters}, }1964]{1964PhRv..136.1224P}
{Peters} P.~C., . 1964.
\newblock {Gravitational Radiation and the Motion of Two Point Masses}.
\newblock {\em Physical Review}, {\bf 136}, 1224--1232.

\bibitem[\protect\citename{Peters and Matthews, }1963]{pm63}
Peters P.~C.,  Matthews J., . 1963.
\newblock ..
\newblock {\em Phys. Rev.}, {\bf 131}, 435.

\bibitem[\protect\citename{Phinney, }1991]{phi91}
Phinney E.~S., . 1991.
\newblock The rate of neutron star binary mergers in the universe - Minimal
  predictions for gravity wave detectors.
\newblock {\em \apjl}, {\bf 380}, L17--L21.

\bibitem[\protect\citename{{Pian} {et~al.}, }2017]{2017Natur.551...67P}
{Pian} E., , {D'Avanzo} P., , {Benetti} S.,    et~al. 2017.
\newblock {Spectroscopic identification of r-process nucleosynthesis in a
  double neutron-star merger}.
\newblock {\em \nat}, {\bf 551}, 67--70.

\bibitem[\protect\citename{{Podsiadlowski} {et~al.},
  }2004]{2004ApJ...612.1044P}
{Podsiadlowski} P., , {Langer} N., , {Poelarends} A.~J.~T.,    et~al. 2004.
\newblock {The Effects of Binary Evolution on the Dynamics of Core Collapse and
  Neutron Star Kicks}.
\newblock {\em \apj}, {\bf 612}, 1044--1051.

\bibitem[\protect\citename{Portegies~Zwart and McMillan, }2000]{pm00}
Portegies~Zwart S.~F.,  McMillan S. L.~W., . 2000.
\newblock Black Hole Mergers in the Universe.
\newblock {\em \apjl}, {\bf 528}, L17--L20.

\bibitem[\protect\citename{Press and Thorne, }1972]{pt72}
Press W.,  Thorne K.~S., . 1972.
\newblock Gravitational-Wave Astronomy.
\newblock {\em \araa}, {\bf 10}, 335--374.

\bibitem[\protect\citename{{Repetto} and {Nelemans},
  }2015]{2015MNRAS.453.3341R}
{Repetto} S.,  {Nelemans} G., . 2015.
\newblock {Constraining the formation of black holes in short-period black hole
  low-mass X-ray binaries}.
\newblock {\em \mnras}, {\bf 453}, 3341--3355.

\bibitem[\protect\citename{{Repetto} {et~al.}, }2012]{2012MNRAS.425.2799R}
{Repetto} S., , {Davies} M.~B.,    {Sigurdsson} S., . 2012.
\newblock {Investigating stellar-mass black hole kicks}.
\newblock {\em \mnras}, {\bf 425}, 2799--2809.

\bibitem[\protect\citename{{Repetto} {et~al.}, }2017]{2017MNRAS.467..298R}
{Repetto} S., , {Igoshev} A.~P.,    {Nelemans} G., . 2017.
\newblock {The Galactic distribution of X-ray binaries and its implications for
  compact object formation and natal kicks}.
\newblock {\em \mnras}, {\bf 467}, 298--310.

\bibitem[\protect\citename{{Rodriguez} {et~al.}, }2016]{2016PhRvD..93h4029R}
{Rodriguez} C.~L., , {Chatterjee} S.,    {Rasio} F.~A., . 2016.
\newblock {Binary black hole mergers from globular clusters: Masses, merger
  rates, and the impact of stellar evolution}.
\newblock {\em \prd}, {\bf 93}(8), 084029.

\bibitem[\protect\citename{{Sathyaprakash} {et~al.},
  }2012]{2012CQGra..29l4013S}
{Sathyaprakash} B., , {Abernathy} M., , {Acernese} F.,    et~al. 2012.
\newblock {Scientific objectives of Einstein Telescope}.
\newblock {\em Classical and Quantum Gravity}, {\bf 29}(12), 124013.

\bibitem[\protect\citename{{Sathyaprakash} and {Schutz},
  }2009]{2009LRR....12....2S}
{Sathyaprakash} B.~S.,  {Schutz} B.~F., . 2009.
\newblock {Physics, Astrophysics and Cosmology with Gravitational Waves}.
\newblock {\em Living Reviews in Relativity}, {\bf 12}, 2.

\bibitem[\protect\citename{{Schutz}, }1989]{1989CQGra...6.1761S}
{Schutz} B.~F., . 1989.
\newblock {Gravitational wave sources and their detectability}.
\newblock {\em Classical and Quantum Gravity}, {\bf 6}, 1761--1780.

\bibitem[\protect\citename{{Sesana}, }2016]{2016PhRvL.116w1102S}
{Sesana} A., . 2016.
\newblock {Prospects for Multiband Gravitational-Wave Astronomy after
  GW150914}.
\newblock {\em Physical Review Letters}, {\bf 116}(23), 231102.

\bibitem[\protect\citename{{Shah} and {Nelemans}, }2014a]{2014ApJ...790..161S}
{Shah} S.,  {Nelemans} G., . 2014a.
\newblock {Constraining Parameters of White-dwarf Binaries Using
  Gravitational-wave and Electromagnetic Observations}.
\newblock {\em \apj}, {\bf 790}, 161.

\bibitem[\protect\citename{{Shah} and {Nelemans}, }2014b]{2014ApJ...791...76S}
{Shah} S.,  {Nelemans} G., . 2014b.
\newblock {Measuring Tides and Binary Parameters from Gravitational Wave Data
  and Eclipsing Timings of Detached White Dwarf Binaries}.
\newblock {\em \apj}, {\bf 791}, 76.

\bibitem[\protect\citename{{Shah} {et~al.}, }2015]{2015JPhCS.610a2003S}
{Shah} S., , {Larson} S.~L.,    {Brown} W., . 2015 (May).
\newblock {Ultra-compact binaries as gravitational wave sources}.
\newblock {Page  012003 of:} {\em Journal of Physics Conference Series}.
\newblock Journal of Physics Conference Series, vol. 610.

\bibitem[\protect\citename{{Sigurdsson} and {Hernquist},
  }1993]{1993Natur.364..423S}
{Sigurdsson} S.,  {Hernquist} L., . 1993.
\newblock {Primordial black holes in globular clusters}.
\newblock {\em \nat}, {\bf 364}, 423--425.

\bibitem[\protect\citename{{Smartt} {et~al.}, }2017]{2017Natur.551...75S}
{Smartt} S.~J., , {Chen} T.-W., , {Jerkstrand} A.,    et~al. 2017.
\newblock {A kilonova as the electromagnetic counterpart to a
  gravitational-wave source}.
\newblock {\em \nat}, {\bf 551}, 75--79.

\bibitem[\protect\citename{{Solheim}, }2010]{2010PASP..122.1133S}
{Solheim} J.-E., . 2010.
\newblock {AM CVn Stars: Status and Challenges}.
\newblock {\em \pasp}, {\bf 122}, 1133--1163.

\bibitem[\protect\citename{{Str\"oer} and {Vecchio}, }2006]{sv06}
{Str\"oer} A.,  {Vecchio} A., . 2006.
\newblock {The LISA verification binaries}.
\newblock {\em Classical and Quantum Gravity}, {\bf 23}, 809.

\bibitem[\protect\citename{{Sukhbold} {et~al.}, }2016]{2016ApJ...821...38S}
{Sukhbold} T., , {Ertl} T., , {Woosley} S.~E., , {Brown} J.~M.,    {Janka}
  H.-T., . 2016.
\newblock {Core-collapse Supernovae from 9 to 120 Solar Masses Based on
  Neutrino-powered Explosions}.
\newblock {\em \apj}, {\bf 821}, 38.

\bibitem[\protect\citename{{Tauris} {et~al.}, }2017]{2017ApJ...846..170T}
{Tauris} T.~M., , {Kramer} M., , {Freire} P.~C.~C.,    et~al. 2017.
\newblock {Formation of Double Neutron Star Systems}.
\newblock {\em \apj}, {\bf 846}, 170.

\bibitem[\protect\citename{Taylor and Weisberg, }1982]{tw82}
Taylor J.~H.,  Weisberg J.~M., . 1982.
\newblock A new test of general relativity - Gravitational radiation and the
  binary pulsar PSR 1913+16.
\newblock {\em \apj}, {\bf 253}, 908--920.

\bibitem[\protect\citename{Thorne, }1997]{tho97}
Thorne K.~S., . 1997.
\newblock Gravitational radiation: a new window onto the universe.
\newblock {\em RvMA}, {\bf 10}, 1--28.

\bibitem[\protect\citename{{Toonen} {et~al.}, }2012]{2012A&A...546A..70T}
{Toonen} S., , {Nelemans} G.,    {Portegies Zwart} S., . 2012.
\newblock {Supernova Type Ia progenitors from merging double white dwarfs.
  Using a new population synthesis model}.
\newblock {\em \aap}, {\bf 546}, A70.

\bibitem[\protect\citename{Tutukov and Yungelson, }1979]{ty79a}
Tutukov A.~V.,  Yungelson L.~R., . 1979.
\newblock "On the influence of emission of gravitational waves on the evolution
  of low-mass close binary stars".
\newblock {\em Acta Astron.}, {\bf 29}, 665--680.

\bibitem[\protect\citename{{Ugliano} {et~al.}, }2012]{2012ApJ...757...69U}
{Ugliano} M., , {Janka} H.-T., , {Marek} A.,    {Arcones} A., . 2012.
\newblock {Progenitor-explosion Connection and Remnant Birth Masses for
  Neutrino-driven Supernovae of Iron-core Progenitors}.
\newblock {\em \apj}, {\bf 757}, 69.

\bibitem[\protect\citename{{van den Heuvel} and {De Loore},
  }1973]{1973A&A....25..387V}
{van den Heuvel} E.~P.~J.,  {De Loore} C., . 1973.
\newblock {The nature of X-ray binaries III. Evolution of massive close
  binaries with one collapsed component - with a possible application to Cygnus
  X-3.}
\newblock {\em \aap}, {\bf 25}, 387--395.

\bibitem[\protect\citename{{van den Heuvel} {et~al.},
  }2017]{2017MNRAS.471.4256V}
{van den Heuvel} E.~P.~J., , {Portegies Zwart} S.~F.,    {de Mink} S.~E., .
  2017.
\newblock {Forming short-period Wolf-Rayet X-ray binaries and double black
  holes through stable mass transfer}.
\newblock {\em \mnras}, {\bf 471}, 4256--4264.

\bibitem[\protect\citename{{Verbunt} {et~al.}, }2017]{2017arXiv170808281V}
{Verbunt} F., , {Igoshev} A.,    {Cator} E., . 2017.
\newblock {The observed velocity distribution of young pulsars}.
\newblock {\em ArXiv e-prints}, Aug.

\bibitem[\protect\citename{Weiss, }1972]{Weiss:1972}
Weiss R., . 1972.
\newblock Electromagnetically coupled broadband gravitational wave antenna.
\newblock {\em Quarterly Progress Report, Research Laboratory of Electronics of
  MIT}, {\bf 105}, 54.

\bibitem[\protect\citename{White and van Paradijs, }1996]{wp96}
White N.~E.,  van Paradijs J., . 1996.
\newblock The galactic distribution of black hole candidates in low-mass X-ray
  binary systems.
\newblock {\em \apj}, {\bf 473}, L25--L29.

\bibitem[\protect\citename{{Willems} {et~al.}, }2005]{whl+04}
{Willems} B., , {Henninger} M., , {Levin} T.,    et~al. 2005.
\newblock {Understanding Compact Object Formation and Natal Kicks I.
  Calculation Methods and the case of GRO J1655-40}.
\newblock {\em \apj}, {\bf 625}, 324.

\end{thebibliography}
\bibliographystyle{cambridgeauthordate}
  
 \copyrightline{} 
 \printindex
    
\end{document}